\documentclass[]{aa}
\usepackage{graphicx}
\usepackage{txfonts}
\begin{document}
   \title{Evolution of Young Brown Dwarf Disks in the Mid-Infrared}

   \author{Michael F. Sterzik,
          \inst{1}
          Ilaria Pascucci,
          \inst{2}
          D\'aniel Apai,
          \inst{2}
          Nicole van der Bliek
          \inst{3}
          \and
          Cornelis P. Dullemond
          \inst{4}
          }

   \offprints{M.F. Sterzik, \email{msterzik@eso.org}}

   \institute{European Southern Observatory, Casilla 19001, Santiago 19, Chile
             \and
             Max-Planck Institute f\"ur Astronomie, K\"onigstuhl 17,
             D-69117 Heidelberg, Germany
             \and
             Cerro Tololo Inter-American Observatory, Casilla 603,
             La Serena, Chile
             \and
             Max-Planck Institute f\"ur Astrophysik, Postfach 1317,
             D-85748 Garching bei M\"unchen, Germany
            }

   \date{Received June 21, 2004; accepted August 30, 2004}

   \abstract{We have imaged two bona-fide brown dwarfs with
   TReCS/GEMINI-S and find mid-infrared excess emission
   that can be explained by optically thick dust disk models. In the
   case of the young ($\approx$2Myr) Cha~H$\alpha$1 we measure
   fluxes at 10.4$\mu$m and 12.3$\mu$m that are fully consistent
   with a standard flared disk model and prominent silicate emission.
   For the $\approx$ 10Myr old
   brown dwarf 2MASS1207-3932 located in the TW Hydrae association we find
   excess emission at 8.7$\mu$m and 10.4$\mu$m  with respect to the photosphere,
   and confirm disk accretion as likely cause of its strong
   activity. Disks around brown dwarfs likely last at least
   as long as their low-mass stellar counterparts in the T-Tauri phase.
   Grain growth, dust settling, and evolution of the
   geometry of brown dwarfs disks may appear on a timescale of 10Myr and
   can be witnessed by observations in the mid-infrared.

   }
  \authorrunning{Michael F. Sterzik et al.}
  \titlerunning{Evolution of Young Brown Dwarf Disks in the Mid-Infrared}
   \maketitle
%
%________________________________________________________________

\section{Introduction}

There is mounting evidence that dusty disks accompany brown dwarfs
(BDs) in their early evolutionary stages similar to circum-stellar
disks around T Tauri stars (TTS). Based on excess emission in the
near-infrared (JHK), Muench et al. (2001) infer that many
substellar objects are initially surrounded by circumstellar
disks. At longer wavelengths the effects of disk emission get more
pronounced compared to the (cool) brown dwarf photosphere.
Jayawardhana et al. (2003) and Liu, Najita \& Tokunaga (2003) find
L' (3.8$\mu$m) excess emission in the majority of substellar
objects in young clusters. The frequency of young circumstellar
disks appears to be similar in the stellar and in the sub-stellar
regime. Physical properties of dust disks around BDs can be better
determined with mid-infrared (MIR) observations. Exploiting the
ISOCAM surveys at 6.7$\mu$m and 14.3$\mu$m in Chameleon (Persi et
al. 2000) and $\rho$-Ophiuchus (Bontemps et al. 2001), Natta \&
Testi (2001) and Natta et al. (2002) tried to explain the spectral
energy distribution (SED) of the associated young  BD population.
They show that the MIR excess emission of these objects is
generally consistent with predictions of simple, passive disk
models. This suggests far-reaching implications like that the
general formation mechanism of stars by core collapse and
formation of an accretion disk extends into the sub-stellar
regime.

Important physical properties of sub-stellar disks remain to be
understood. For example, the assumption of hydrostatic equilibrium
in the vertical disk direction -- reasonably accurate for many T
Tauri disks -- implies a flared geometry, and the presence of an
optically thin surface layer that is heated by direct illumination
from the central source (Chiang \& Goldreich 1997; Chiang et al.
2001; Natta et al. 2001). In these upper layers a prominent
emission signature of silicates is expected around 10$\mu$m, as
observed for many disks around Herbig Ae/Be (Meeus et al. 2001)
and TTS (Meeus et al. 2003; Pryzgodda et al. 2003). However, the
majority of disk models that actually fit SEDs of young BDs
apparently require a flat geometry (Natta et al. 2002).
Alternatively, or in addition to geometrical effects, also grain
growth can weaken, or even extinguish the 10$\mu$m signature
(Bouwman et al. 2001). Unfortunately, the poor accuracy and
wavelength sampling of the ISO photometry hamper to draw
unambiguous conclusions.

New, sensitive, mid-infrared instruments at large telescopes allow
us to probe the 10$\mu$m wavelength range of the SEDs of young BDs
in much greater detail and with higher accuracy. The first
ground-based MIR detection of a disk around the young BD candidate
ChaH$\alpha$2 unequivocally demonstrates the absence of a silicate
emission feature in its disk (Apai et al. 2002). Both, flared and
flat disk models, are consistent to explain the SED of three young
brown dwarfs GY5, GY11 and GY310 (Mohanty et al. 2004). Apai et
al. (2004) present first evidence for grain growth and dust
settling within the disk of the young brown dwarf CFHT BD Tau 4,
the best characterized BD disk up to date (Pascucci et al. 2003;
Klein et al. 2003).

In this paper we report high-quality MIR photometry of two
bona-fide BDs, namely ChaH$\alpha$1 and 2MASS1207-3932 (in the
following 2M1207), obtained with the Thermal-Region Camera
Spectrograph T-ReCS mounted at Gemini-South. A disk around
ChaH$\alpha$1 has already been inferred from ISOCAM observations
(Natta \& Testi 2001). With the help of our new photometric values
we refine its disk model. For 2M1207 we present the first direct
evidence for the existence of circum-sub-stellar material around
an  approximately 10Myr old object. 2M1207 has been identified as
BD by optical spectroscopy (Gizis 2002), and is related to the TW
Hydrae association (TWA), a loose group of $\sim$~10Myr old T
Tauri stars at an approximate distance of 60~pc to the sun (Webb
et al. 1999). The lack of a measured $K-L'$ excess did not
indicate the presence of any circum-sub-stellar material
(Jayawardhana et al. 2003),  but strong emission in H$\alpha$ and
HeI suggested indirectly that this object may be a (weak) accretor
(Mohanty, Jayawardhana \& Barrado y Navascues 2003; in the
following MJBN). Our new measurements -- to our knowledge among
the most sensitive ground-based mid-infrared photometry ever
reported -- demonstrate the persistence of dust around older BDs,
on timescales at least as long as for low-mass stars.

\section{Observations and Data Reduction}

\begin{figure}
\centering
\includegraphics[bb= 20 20 412 312, width=7cm, clip]{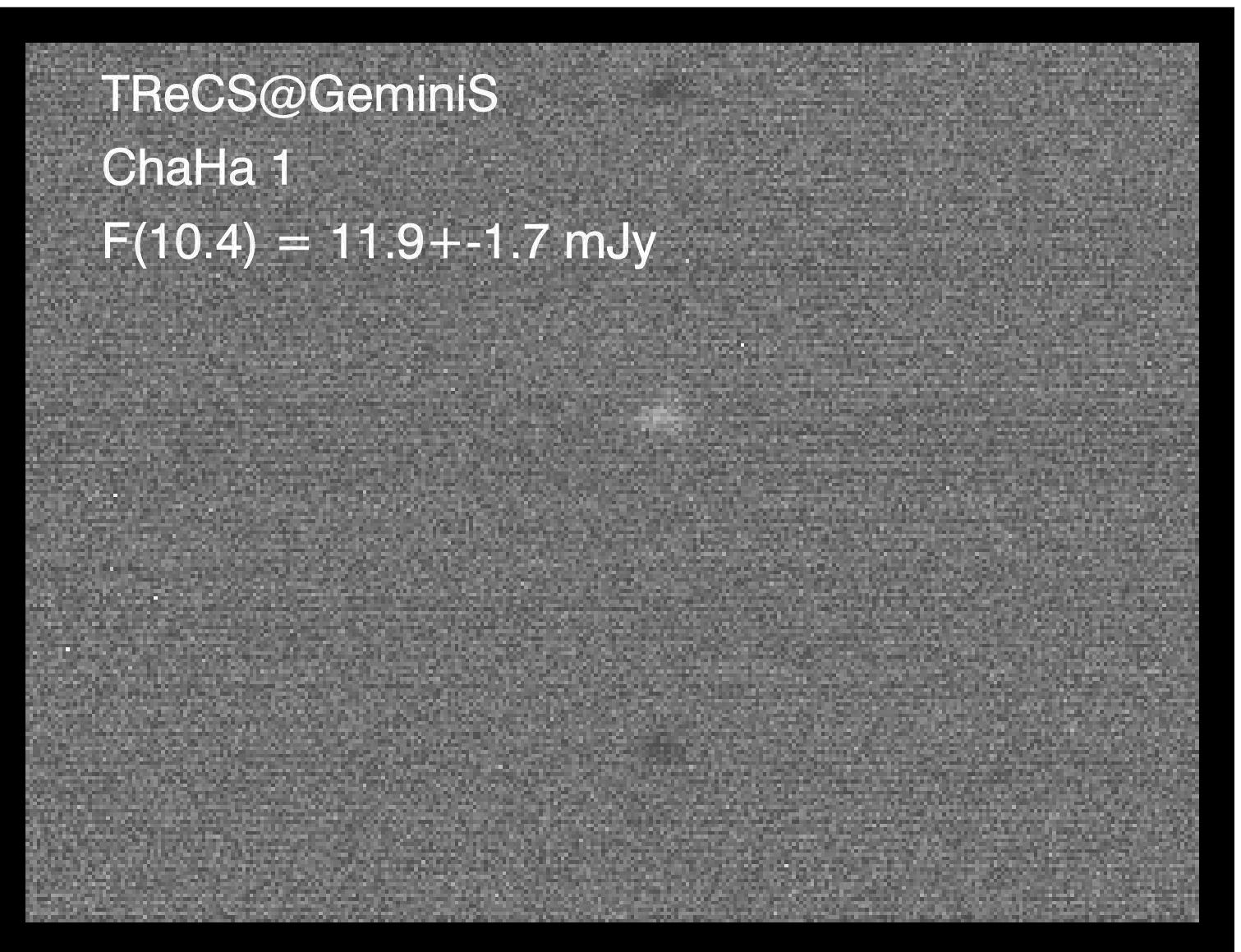}
\caption{TReCS image of ChaH$\alpha$1 at 10.4$\mu m$. Note the
smearing due to imperfect tracking.} \label{chaha1}
\end{figure}

\begin{figure}
\centering
\includegraphics[bb= 20 20 412 312,width=7cm, clip]{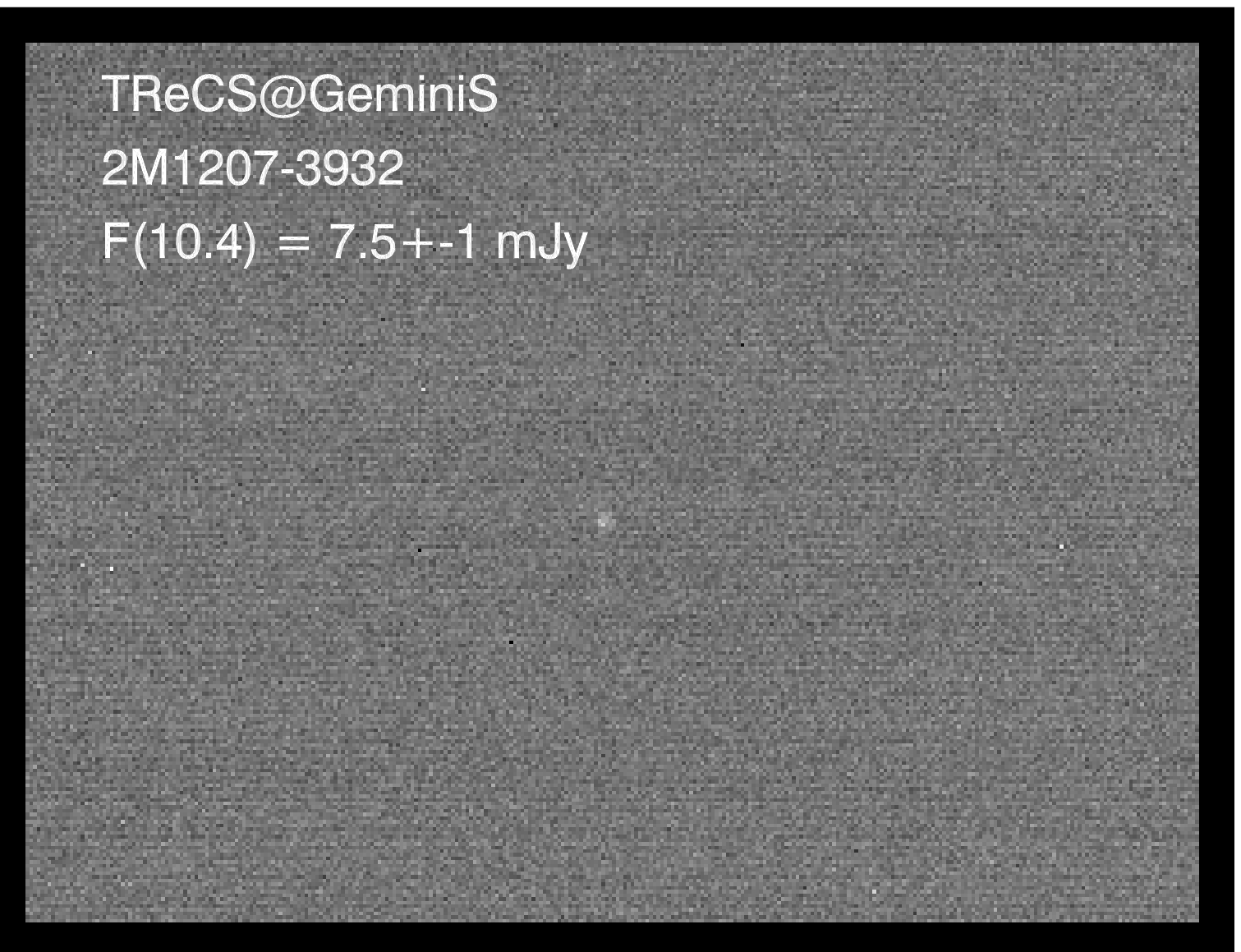}
\caption{TReCS image of 2M1207-3932 at 10.4$\mu m$. Chop and nod
amplitudes were larger then the field-of view ($\approx 29" \times
22"$)} \label{2m1207}
\end{figure}

The imaging observations were performed in service mode with TReCS
(Telesco et al. 1998) mounted at Gemini-South on Jan. 2, 2004
(ChaH$\alpha$1) and on Jan. 27, 2004 (2MASS1207-3932) under clear
weather conditions. Its pixel scale of 0.09" results in a field of
view of $\approx 29" \times 22"$. The same throw amplitudes, but
opposite directions, were used for chopping and nodding. The
observations on Jan. 2 were obtained with throw amplitudes of 10
arcsec, and the positive and two negative beams are all located in
the field of view. The source appears only in the final, co-added
frame, because slight tracking errors of the telescope introduced
some image smearing of about 2-3 pixels. A more accurate
correction of the tracking errors was not possible due to the
faintness of ChaH$\alpha$1, and a "shift and add" co-add of
individual chopping pairs was not possible (in contrast to the
case of BD-Tau4 discussed in Apai et al. 2004). Fig.~\ref{chaha1}
shows the final frame with the detection. 2M1207 was observed with
15 arcsec throw amplitudes, and only the overlay of the two
positive beams can be extracted in the final image. No tracking
errors were evident, and the target appears point-like in the
final image (Fig.~\ref{2m1207}). A summary of the observing log,
including total integration times, is given in Table~\ref{obslog}.

\begin{table}
      \caption[]{Log of the TReCS observations.}
         \label{obslog}

         \begin{tabular}{llllll}
            \hline
            \hline
            \noalign{\smallskip}
            RJD  &  Object &  Filter &  Airmass & Int[s]  &  Calibrator   \\
            \noalign{\smallskip}
            \hline
            \noalign{\smallskip}
            53006.74 & ChaH$\alpha$1 & Si1-7.7 & 1.6 & 782 &
            HD92305 \\
            53006.70 & ChaH$\alpha$1 & Si4-10.4 & 1.6 & 912 &
            HD92305 \\
            53006.79 & ChaH$\alpha$1 & Si6-12.3 & 1.5 & 1216 &
            HD92305 \\
            53031.84 & 2M1207 & Si1-7.7 & 1.0 & 304 &
            HD110458 \\
            53031.85 & 2M1207 & Si2-8.7 & 1.0 & 608 &
            HD110458 \\
            53031.81 & 2M1207 & Si4-10.4 & 1.0 & 782 &
            HD110458 \\
            \noalign{\smallskip}
            \hline
         \end{tabular}
   \end{table}

Final co-added images were obtained from TReCS pipeline processed
images and from manual merging of individual frames, with
consistent results. Source count-rates were extracted with
standard aperture photometry. Curve-of-growth methods were applied
to find the optimal aperture radius that maximized the
signal-to-noise ratio of the extracted source. That radius was
then applied for the calibration source to determine the
count-rate to flux conversion factor. Error estimates are based on
the formal error of the background noise weighted over the source
extraction region. When positive and negative beams were
available, the fluxes were found to be consistent with these
formal error estimates. The stellar parameters assumed, and
mid-infrared fluxes observed are summarized in
Table~\ref{results}.

\begin{table*}
      \caption[]{Stellar parameters and mid-infrared fluxes.
      7.7$\mu $m upper limits, and 8.7$\mu $m, 10.4$\mu $m, 12.3$\mu $m measurements are from this work.}
         \label{results}

         \begin{tabular}{lllllllllllll}
            \hline
            \hline
            \noalign{\smallskip}
            Object  &  SpT &  T$_{eff}$ & L$_{bol}$ & M  & age & 3.8$\mu m^a$ & 6.7$\mu m$ & 7.7$\mu m^d$
            & 8.7$\mu m$ & 10.4$\mu m$ & 12.3$\mu m$ &14.3$\mu m$    \\
            &      &  [K]      & [L$_\odot$] & [M$_\odot$] & [Myr] & [mJy]
            & [mJy] & [mJy] & [mJy] & [mJy] & [mJy] & [mJy] \\
            \noalign{\smallskip}
            \hline
            \noalign{\smallskip}
            ChaH$\alpha$1 & M7.5$^b$ & 2800     & 0.014     & 0.05$^b$  & 2$^{a,b}$ & 7$\pm$1 & 6.7$\pm$1.8& $<$11$^e$ & - & 11.9$\pm$1.7& 11.9$\pm$0.7& 11.8$\pm$1.5\\
            2M1207        & M8$^c$   & 2600     & 0.0035    & 0.03      & 10$^a$ & 7$\pm$1 &  - & $<$8 & 5.6$\pm$1 & 7.5$\pm$1& - & - \\
           \noalign{\smallskip}
            \hline
         \end{tabular}
         \begin{list}{}{}
\item[$^{\mathrm{a}}$] Jayawardhana et al. (2003)
\item[$^{\mathrm{b}}$] Comeron, Neuh\"auser \& Kaas (2000) ($A_I=0.11$)
\item[$^{\mathrm{c}}$] Gizis (2002)
\item[$^{\mathrm{d}}$] upper limit refer to a 5$\sigma$ detection
threshold
\item[$^{\mathrm{e}}$] smearing assumed (FWHM=1arcsec)
\end{list}
   \end{table*}

%%Ilaria, pls explain: JHK fitting? assumptions, extinction...})

Given the effective temperatures and bolometric luminosities
indicated in Table~\ref{results}, the purely photospheric
contribution to the 10.4$\mu$m flux density is about 1mJy for
ChaH$\alpha$1 and 1.5mJy for 2M1207. This estimate is based on BD
atmospheric models  from the Lyon group (e.g. Allard et al., 2000,
2001). For the following, we use their published tables that
include full dust treatment (both condensation and opacities) and
use the TiO and H$_2$O line lists from Nasa AMES. We refer to a
surface gravity of $\log(g)=3.5$ and solar metallicity. The
atmospheric fluxes in the N-band in these models do not vary by
more then 1\% from a blackbody flux assuming the same $T_{eff}$
(in contrast to the near-IR, where the deviations can be huge).
Hence, both our measurements indicate significant excess that
cannot come from a BD photosphere alone. An excess criterium based
only on NIR photometry (like $E(K-L')_0$ used e.g. in Jayawardhana
et al. 2003) has very limited value to signal the presence of
circum-sub-stellar material, as it appears only marginally
indicative (for ChaH$\alpha$1) or even fails (for 2M1207).

\section{Disk Models}

In this section we explain the main features of the SED using the
new TReCS photometry (combined with ISO photometry, if available)
with models of BD disks. The models are based on the Chiang \&
Goldreich (1997) approximations, with modifications from
Dullemond, Dominik \& Natta (2001) for flux conservation. The
general methodology and application to BD disks is explained in
Pascucci et al. (2003). In view of the uncertainties of the exact
shape of the SED over a wider wavelength range due to the
knowledge of only very few data points, we do not attempt to
fine-tune the disk models with respect to variations of input
parameters like disk inner and outer radii, or inclination. We
merely want to concentrate the discussion on those parameters that
have a most notable impact on the shape of the SED around
10$\mu$m. These are mainly the assumed disk geometry, and the
dominant size and composition of the dust particle distribution.
In the following, we assume standard astronomical silicates with
opacities from Draine \& Lee (1984). As we discuss only passive,
re-processing disks, the assumed parameters of the BD mass and its
luminosity are important. We directly use, or derive them from
published values as indicated.

\subsection{ChaH$\alpha$1}

%{\bf Ilaria: why do we not simply assume the ChaHa1 parameters
%given in Comeron, Neu\"auser \& Kaas (2000): age = 2Myr, mass =
%0.04, Lbol=0.011, Teff=2770K). I think this would simplify the
%discussion. Alternatively pls explain, why we prefer to use
%different values.}

\begin{figure}[t]
\centering
\includegraphics[width=7cm, angle=90]{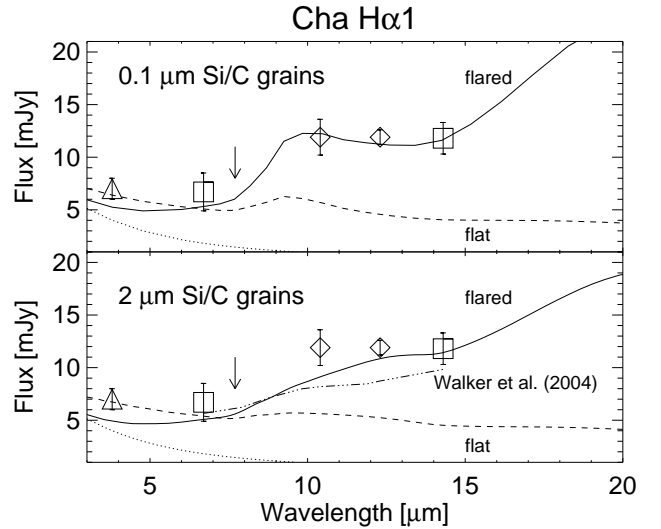}
\caption{SED and comparison of disk models for ChaH$\alpha$1.
Measurements are diamonds, with associated error bars. The upper
panels refers to small dust sizes (0.1$\mu$m), the lower panel to
larger grains (2$\mu$m). Flared (full line) and flat (dashed line)
model predictions are compared. The contribution of the BD
photosphere is approximated by a black-body model. The
dashed-dotted lines refers to the disk model by Walker et al.
(2004). Our T-ReCS measurements are shown with diamonds, while the
ISOCAM data are indicated with boxes and the L' point with a
triangle, respectively.} \label{cha_mod}
\end{figure}

According to Comeron, Neuh\"auser \& Kaas (2000), the spectral
type of ChaH$\alpha$1 is M7.5, and they estimate $T_{eff}=2770$K
and $L=0.011L_{\odot}$. Using the dusty BD model grids from Allard
et al. (2000, 2001), we find that the available near-IR photometry
for this object is in much better agreement with an atmospheric
model for a $\approx$ 1Myr old BD with $T_{eff}=2800$K and
$L=0.014L_{\odot}$.  We prefer these values for the BD
photospheric parameters, which are anyway consistent with the
Comeron et al. estimates, given their large error bars.
Fig.~\ref{cha_mod} summarizes the results of our disk modeling in
the MIR regime. The upper panel refers to opacities of
astronomical (amorphous) silicates of 0.1$\mu$m size, while the
lower panel displays the results of large, 2$\mu$m sized grains of
the same composition. Full lines indicate the SED expected for a
flared model, in which the vertical scale height is derived from
local hydrostatical equilibrium at each disk radius. The dashed
lines refer to the SED in a flat geometrical configuration.
%({\bf is this correct??}).
Contributions of the photosphere are indicated as dotted lines.
The dashed-dotted line indicates the results from Walker et al.
(2004). We will refer to it later.

We note that the L' photometry from Jayawardhana et al. (2003)
already indicates a significant excess, and favors - according to
our modeling - a flattish disk geometry. However, the superb
quality of our new TReCS measurements, especially at 10.4$\mu$m,
constrains the disk configuration to a standard, flared disk
model. Standard, small, ISM grains can explain the silicate
emission feature very well, although larger grains cannot be
excluded.

\subsection{2MASS1207-3932}

Following Gizis (2002), the spectral type of 2M1207 is M8, which
translates to $T_{eff}=2600$K according to the calibration of
Luhman (1999). If we assume an age of 10Myr (see below), and scale
the available JHK photometry to a distance of 70pc, we find an
approximate bolometric luminosity of 0.0035L$_{\odot}$ and a mass
of $0.003 M_\odot$ for this BD from the evolutionary tracks of
Allard et al. (2000, 2001). These values are consistent with those
estimated in MJBN.

The interpretation of the SED for this BD is not unique. ISO
observations are not available, and the two TReCS measurements,
together with the upper limit at 7.7$\mu$m  and the L$^\prime$
measurement clearly exclude a standard flared disk model, with ISM
grain sizes (0.1$\mu$m). Either a flat disk, and/or somewhat
bigger grains (2$\mu$m) are compatible with the photometry. But
large grains (up to 5$\mu$m) only seem to be possible in a flared
configuration viewed from a high inclination angle (50deg). The
flat geometrical configuration is, on the other hand, more robust
with respect to inclination angle variations, because shadowing
effects are less pronounced.

Recently, Chauvin et al. (2004) report the detection of a giant
planet candidate close to 2M1207. From near-IR spectroscopy they
derive a spectral type of L5-L9.5, and a likely temperature of
$T_{eff} = 1250 \pm 200$K for this object, only 0.8 arcsec
separated from the primary. Although it should be possible to
spatially resolve it with a 8m class telescope, we do not have any
evidence for this object in any of our images (see
Fig.~\ref{2m1207}). But we note that the expected photospheric
flux in the N-band is only around 0.2mJy, assuming a simple
black-body scaled to their $K$-band measurement. We therefore
conclude that the excess flux we see is originating only from the
disk material around 2M1207, and is not contaminated by the
putative companion.

\begin{figure}[t]
\centering
\includegraphics[width=7cm, angle=90]{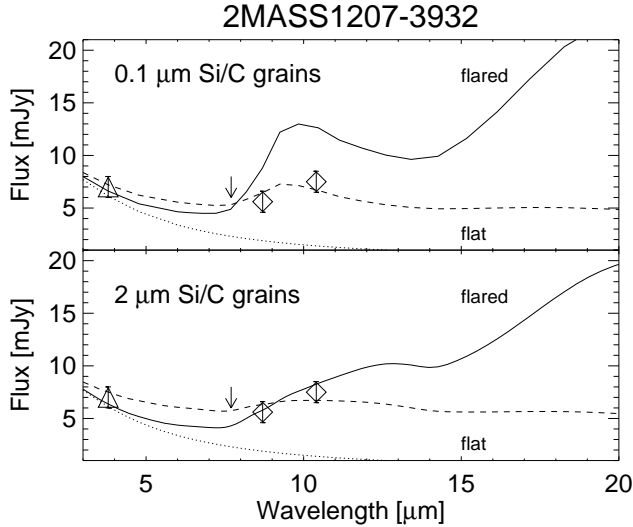}
\caption{SED and comparison of disk models for 2M1207.}
\label{2m_mod}
\end{figure}

\section{Discussion}

Even with the new and improved MIR photometry an unambiguous
interpretation remains difficult, because grain sizes and dust
spatial distribution are coupled. For both objects, flared models
can in principle explain the observed MIR fluxes. A flared disk
containing small grains heated in the upper surface layers is the
most likely explanation for the pronounced silicate feature in
ChaH$\alpha$1. A flat geometry, and substantial grain-growth are
excluded for ChaH$\alpha$1. In the case of 2M1207 we find that a
flared disk requires a large inclination {\em and} grains with
sizes of a few $\mu$m to fit the two 10$\mu$m points. Equally
well, a flat  disk composed of small, or moderately large grains
up to $\approx 2\mu$m can account for the observations.

We will now try to put these observations in an evolutionary
scenario.

\subsection{Grain Growth and Age Evolution}

TWA membership of 2M1207 has been deducted both from proper motion
(Gizis 2002), and radial velocities (MJBN). The age of the
association itself is fairly well constrained: 8.3Myrs from
dynamical age determination (Makaraov \& Fabricius 2001),
$\approx$ 10Myrs from isochrone fitting of its stellar population
(Webb et al. 1999), and a similar age from Li abundance
determination for late-type members (e.g. Soderblom et al. 1998).
However, the presence of optically thick circum-stellar disks in
many of the prominent stellar members of TWA (TW-Hya, HD98800B,
Hen3-600A, HR4796; see Jayawardhana et al. 1999) and signatures of
active accretion like in the classical TTS TW-Hya (Rucinski \&
Krautter 1983; Muzerolle et al. 2000) has always challenged a
consistent evolutionary picture for this group of stars. Gap
formation and significant dust evolution are observed in all four
stellar systems mentioned above, and the detailed analysis of
low-resolution N-band spectra allows to decompose the 10$\mu$m
feature in distinct dust species: TW-Hya exhibits broad silicate
emission, indicative for amorphous, large (2$\mu$m) olivine grains
(Weinberger et al. 2002).  HD98800B has highly processed dust,
dominated by large amorphous olivine and crystalline forsterite
(Sch\"utz, Meeus \& Sterzik 2004). And also the spectrum of
Hen3-600A features a rich mixture of crystalline silicate
components, and even SiO$_2$ (Honda et al. 2003).

In other words, {\em all} circum-stellar disks in TWA show signs
of dust processing, and are in particular incompatible with a
primordial dust size distribution like the ISM. Hence, the dust
observed in these disks must have evolved on a 10Myr timescale
since star formation. If the disk evolution is qualitatively
similar around stars and sub-stellar objects, then one might
expect similar dust properties also in circum-sub-stellar disks.
The indication of larger grains in 2M1207 from our modeling is
therefore fully consistent with this notion.

The case of ChaH$\alpha$1 fits in this evolutionary scenario. Not
much dust processing and/or growth has occurred, and one reason
for this could well be its relatively young age (2~Myrs, or even
$<0.1$~Myrs, Gomez \& Mardones 2003). Small grains are well-mixed
throughout the vertical structure, which is itself in
hydrostatical equilibrium, and illuminated by the central source.

\subsection{Comparison with other BDs}

Beside the two BDs discussed in this paper, the relation of disk
geometry and grain size distribution has been investigated only in
very few other BDs based on (ground-based) MIR photometry that
sample the silicate feature well enough.

Mohanty et al. (2004) discuss three BDs in the young ($<1$Myr)
$\rho$-Ophiuchus star-forming region. For two of these objects (GY
310 and GY 11) they favor a standard, flared disk geometry, while
a flat geometry cannot be ruled out for their third object (GY 5).
Their data do not allow to draw strong constraints on the grain
sizes, but small grains appear to better fit the silicate feature
- at least in GY 310. In summary, these three, very young BD, do
not show evidence for significant grain and disk evolution,
consistent with the notion that grain growth occurs at later
stages, during disk evolution.

But this simple, evolutionary, picture might be more difficult to
reconcile with the finding that other brown dwarf disks of similar
age contain non-primordial dust, and larger grains then in the
ISM. Good evidence for grain-growth is, e.g., presented in Apai et
al. (2004) for the $\approx$1Myr old brown dwarf BD~Tau~4.
Although a precise age determination always suffers from
notoriously uncertain assumption about sub-stellar evolutionary
tracks and unknown distances, a much older age seems to be
excluded because the object is already over-luminous for that
particular isochrone (Mart\'in et al. 2001). Nevertheless, this BD
disk deviates from an equilibrated flared disk geometry, and
rather suggests that a significant amount of grain growth and dust
settling has already occurred in its short lifetime. Also in the
case of ChaH$\alpha$2, likely of similar age as ChaH$\alpha$1,
only a flat disk geometry (Apai et al. 2002), and/or a somewhat
larger grain size distribution (Walker et al. 2004) can explain
the observed SED. Both possibilities are indicative for some
dust/disk evolution that has not -- yet -- happened in
ChaH$\alpha$1.

When, and how, does grain growth in circum(-sub)-stellar disks
occur?

\subsection{Primordial Grain Growth}

There is some observational evidence that larger grains (and
therefore opacities having a shallow wavelength dependency) are
observable very early in disks around T Tauri stars (see e.g. Wood
et al. 2002). It has been suggested that grain growth might
therefore happen already during the collapse phase, or very early
in the formation phase of the circum-stellar disk (see, e.g.
Suttner \& Yorke 2001). Based on this assumption, Walker et al.
(2004) model the SED of BD disks, too, and find reasonable
agreement with those 12 BD disks that have ISOCAM MIR photometry.
But not all of their model predictions are compatible with our
observations. The Walker et al. (2004) model (indicated by a
dot-dashed line in Fig.~\ref{cha_mod}) fails, e.g., for
ChaH$\alpha$1. Our new TReCS measurements rule out the possibility
that the observed opacities (and therefore the grain sizes)
deviate significantly from ISM properties. This means that at
least for this objects grain growth, and dust processing, is not
(yet) dominating the dust grain population as observed in the MIR.

\subsection{The Role of Multiplicity}

The co-existence of dust disks with vastly different grain
properties has also been found in samples of classical TTS disks
(Pryzgodda et al. 2003). Meeus et al. 2003, in particular, analyse
the dust properties in a co-eval sample of TTS of similar (late)
spectral type in Chameleon.  A detailed mineralogical analysis of
the N-band spectra of each individual object requires a different
dust composition and size distribution, ranging from primordial
ISM to highly processed dust. They note that the object that shows
the highest degree of dust processing is in a close binary star,
while the apparently unprocessed dust disk belongs to a single
star. Without drawing strong conclusions they speculate that the
presence of a binary might influence and help dust processing in
disks. We might witness a similar trend in the observed BD disks
in Chameleon. ChaH$\alpha$2 has been reported to be a close (0.2")
binary (Neuh\"aeuser et al. 2002), and its disk exhibits signs of
significant dust processing and grain growth, in contrast to
ChaH$\alpha$1, apparently a single BD (Joergens \& Guenther 2001).

There is no detailed model that explains the influence of a binary
potential on the dust disk dynamics. But it seems plausible that a
perturbing force acting on a circum-sub-stellar disk can induce
enhanced collision, and therefore increase growth rates of dust
particles in its disk (see, e.g. Dubrulle, Morfill \& Sterzik
1995).

In this light, the overluminosity of BD~Tau~4 in the HR diagram,
together with its dust properties, might be a consequence of a
close companion. Interestingly, Itoh, Tamura \& Nakajima (1999)
find a faint companion candidate around this object, though at
rather wide separation ($\approx 4.2"$), and its physical relation
to BD~Tau~4 is not established.

\section{Summary}

We report new high-quality MIR photometry from TReCS of two
bona-fide brown dwarfs, and find excess emission in both cases,
indicative for optical thick dust disks. We have investigated the
influence of diverse disk geometries (flared versus flat) and of
different grain size populations (0.1$\mu$m versus 2$\mu$m) on the
MIR-SED of these brown dwarfs. In particular, we find:
\begin{itemize}
\item The MIR-SED of the $\approx$ 2Myr old ChaH$\alpha$1 is fully
consistent with a standard flared disk model, and primordial (ISM)
grain size properties.
\item The MIR-SED of the $\approx$ 10Myr old 2M1207 can be better
explained by the presence of larger dust, and/or a flat disk
geometry.
\item The evolutionary timescale of dust in brown dwarfs appears
to  be similar as in TTS.
\item Similar to TTS, rapid grain growth and dust processing early in the
BD disk evolution might be related to the presence of a close
companion.
\end{itemize}

\begin{acknowledgements}
      We thank C. Walker for providing her model fluxes of
      ChaH$\alpha$1 in electronic form.
      Based on observations obtained at the Gemini Observatory,
      which is operated by the Association of Universities for Research in Astronomy,
      Inc., under a cooperative agreement with the NSF on behalf of the Gemini partnership:
      the National Science Foundation (United States),
      the Particle Physics and Astronomy Research Council (United Kingdom),
      the National Research Council (Canada), CONICYT (Chile),
      the Australian Research Council (Australia), CNPq (Brazil), and CONICET (Argentina).
      We are indept to the TRecS/Gemini teams providing us excellent data from
      service observations."
\end{acknowledgements}

%%%%%%%%%%%%%%%%%%%%%%%%%%%%%%%%%%%%%%%%%%%%%%%%%%%%%

\end{document}